# Improving Deep Attractor Network by BGRU and GMM for Speech Separation

*Rawad Melhem* *, Assef Jafar and Riad Hamadeh*

(*Higher Institute for Applied Sciences and Technology, Damascus, Syria*)

**Abstract**: Deep Attractor Network (DANet) is the state-of-the-art technique in speech separation field, which uses Bidirectional Long Short-Term Memory (BLSTM), but the complexity of the DANet model is very high. In this paper, a simplified and powerful DANet model is proposed using Bidirectional Gated neural network (BGRU) instead of BLSTM. The Gaussian Mixture Model (GMM) other than the k-means was applied in DANet as a clustering algorithm to reduce the complexity and increase the learning speed and accuracy. The metrics used in this paper are Signal to Distortion Ratio (SDR), Signal to Interference Ratio (SIR), Signal to Artifact Ratio (SAR), and Perceptual Evaluation Speech Quality (PESQ) score. Two speaker mixture datasets from TIMIT corpus were prepared to evaluate the proposed model, and the system achieved 12.3 dB and 2.94 for SDR and PESQ scores respectively, which were better than the original DANet model. Other improvements were 20.7% and 17.9% in the number of parameters and time training respectively. The model was applied on mixed Arabic speech signals and the results were better than that in English.

**Keywords**: attractor network; speech separation; gated recurrent units

**CLC number**: TN912.3　　**Document code**: A　　**Article ID**: 1005-9113(2021)03-0090-07

## 0　Introduction

Isolating each speech signal from a mixture in noisy environment is an easy task for human but a difficult work for machine. The problem is called "cocktail party", and its solution is useful in various applications, such as automatic meeting transcription, automatic caption for Audio/Video recordings (e.g., YouTube), applications that need human-machine interaction (e.g., Internet of things (IoT)), and advanced hearing aids.

Cocktail party was formalized by Cherry E C in 1953[1], and many solutions were proposed. Most of the proposed solutions tried to mimic actions of human ears by filtering or extracting auditory properties from mixture, and then grouping T-F bins of the same speaker. These methods belong to Computational Auditory Speech Analysis (CASA)[2], but CASA methods are not enough for analysis. Another approach for multi-speaker separation is non-negative Matrix Factorization (NMF)[3], which decomposes spectrogram matrix into two matrices (templates and activations). By using activations and non-negative dictionaries, the separated signal can be approximated. Statistical methods were also utilized as solutions for speech separation, such as Independent Component Analysis (ICA)[4], which assumes that speech signal and interference signal are statistically independent, so the separation can be conducted by maximizing the independence. However, ICA only works for overdetermined environment, while cocktail party is an issue in underdetermined environment[5].

Deep learning performs significantly better than other methods, which has been applied in speech enhancement and separation[6-9] as well as in music separation[10-11]. However, deep learning has two obstacles[12], i.e., fixed number of outputs and permutation of sources. Training neural network to separate (n) signals does not work for any number that differs from (n), thus resulting in fixed number of outputs. Permutation of sources occurs due to the order of sources at the outputs of network. Training the network on two different orders of sources will increase the training error and lead to convergence problem. The difficulty of "permutation of sources" can be solved through Permutation Invariant Training （PIT）[13] by calculating the error between separated signal and the targets and choosing the target corresponding to the minimum error. However, PIT can only remove the permutation obstacle, while the problem of fixed number of outputs remains. Deep clustering （DC）[14] outperforms PIT in solving the problems of "fixed number of outputs" and "permutation of sources". The main idea of DC is to produce new space of embeddings, which has desirable features that can make speaker segmentation much easier. Each T-F bin of spectrogram corresponds to an embedding vector, so the embedding space is dense and able to show the T-F bins in a separable way. Clustering algorithm is applied to the embeddings to get each speaker, so it is feasible to determine the number of clusters to vary the outputs of the network. In this way, DC can solve the problem of fixed number of outputs. Loss function of DC is Frobenius norm between affinity matrix of embedding and affinity matrix of target binary mask, so the permutation problem disappears since the affinity matrices are not affected by the order. The trouble of DC is that it does not represent end-to-end system, because mask generation is done separately after the neural network[12]. Deep Attractor Network （DANet）[12] depends on DC algorithm, but after extracting the embeddings, central points of each speaker's embedding will be created, which are called "attractors". By calculating the distance between each embedding vector and the attractors, the mask for each speaker will be generated. DANet uses reconstruction error by comparing between reconstructed and ideal signals, so it provides end-to-end system. The disadvantage of DANet is its complexity, which takes a long time in training, and relatively long time in estimating the masks. Gated Recurrent Units （GRU）[15] is a new version of recurrent neural network （RNN）, which has better results than LSTM in many cases [16]. In this paper, a new version of DANet is proposed, which is less complex and more accurate. Embeddings were created by Bidirectional GRU instead of BLSTM, which can make the model less complex. Therefore, the neural network can be trained on normal workstation. Gaussian Mixture Model （GMM） clustering algorithm was employed, instead of k-means, for a more accurate model.

The rest of the paper is organized as follows. Speech separation problem is introduced in Section 1.

In Section 2, the proposed system is explained. The experimental results are discussed in Section 3.

## 1 Single Channel Speech Separation Problem

Single channel speech separation problem is defined as follows：

Estimate the signals $s_i(t)$, $i = 1,2,\cdots,N$, given only signal $\boldsymbol{x}(t)$, where $\boldsymbol{x}(t)$ is the mixture of $N$ speech signals $\boldsymbol{x}(t) = \sum_{i=1}^{N} s_i(t)$.

To write $\boldsymbol{x}(t)$ in time-frequency （T-F） domain, Short-time Fourier Transform （STFT） is calculated as

$$X(f,t) = \sum_{i=1}^{N} S_i(f,t) \tag{1}$$

where $X$, $S_i$ are the Fourier transform for $\boldsymbol{x}$ and $s_i$, respectively.

## 2 The Proposed Model

The proposed method is very similar to DANet[12], except that the hidden Layers are BGRU rather than BLSTM, and GMM is the alternative to the k-means algorithm.

### 2.1 GRU

Fig. 1 shows the architecture of LSTM and GRU[17].

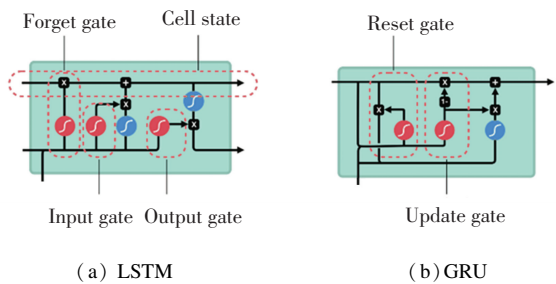

（a）LSTM      （b）GRU

**Fig.1　Architecture of LSTM and GRU**

RNN can use hidden state to process sequences of inputs, and there are three categories of RNN：

1）Vanilla RNN：It is the simplest type of RNN, which has problems when training "vanishing/ exploding gradient".

2）LSTM：It first appeared in 1997, and has three gates, i.e., input, forget, and output. The method can solve the vanishing/exploding gradient but with high complexity.

3）GRU：It was found in 2014, and is considered the simplest version of LSTM, which replaces forget and input gates with update gate.

LSTM has achieved satisfactory results in speech processing tasks，but the complexity remains to be a trouble，so the network needs more data to learn. Recently，GRU has become a strong competitor of LSTM，and in some cases both of them have nearly the same result，but GRU outperforms the standard LSTM in most times[16]. Merging input and forget gates in one-gate results in combining hidden state with cell state，which is one of the reasons for the superiority of GRU[18].

## 2.2 GMM

k-means is a special case of GMM，while it has restrictions that it is only suitable when clusters are spherical. The biggest limitation of k-means is probably that each cluster has the same diagonal covariance matrix. GMM is more flexible and almost more accurate than k-means. In this paper，it is proposed that each cluster resulting from GMM has its own general covariance matrix.

Fig.2 shows the flow chart of the whole method. The steps can be summarized as preprocessing，training phase，and testing phase.

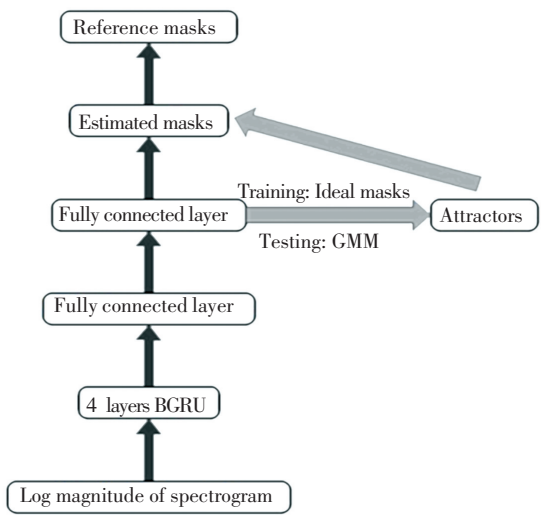

**Fig.2   The proposed model using BGRU and GMM**

### 2.2.1   Preprocessing

1）Calculate the magnitude of spectrogram of the mixture using STFT as flatten vector $X \in \mathbf{R}^{1 \times FT}$ to be the input for the neural network.

2）Build ideal binary mask for each speaker using Weiner-filter like mask as

$$\mathrm{WFM}_{i,ft} = \frac{|s_{i,ft}|^2}{\sum_{j=1}^{N} |s_{j,ft}|^2} \qquad (2)$$

$$\boldsymbol{m}_i = \begin{cases} 1, & \mathrm{WFM}_{i,ft} > \tau \\ 0, & \mathrm{otherwise} \end{cases} \qquad (3)$$

where $\boldsymbol{m}_i \in \mathbf{R}^{1 \times FT}$.

Choose $\tau = 0.5$ as a threshold. Ideal masks will be used in training phase only.

### 2.2.2   Training phase

1）Generate the embedding space $V$ using four BGRU layers and fully connected layer，where each T-F bin of magnitude of spectrogram maps $K$-dimensional vector.

$$V = f(X)，\qquad V \in \mathbf{R}^{K \times FT} \qquad (4)$$

The fully connected layer is used to represent the spectrogram into the dense embedding space.

2）Form attractor $\boldsymbol{a}_i \in \mathbf{R}^{1 \times K}$ for each cluster by calculating the weighted average of embeddings as

$$\boldsymbol{a}_i = \frac{\boldsymbol{m}_i \cdot V^{\mathrm{T}}}{\sum_{f,t} \boldsymbol{m}_i}，i = 1,2,\cdots,N \qquad (5)$$

3）Measure the distance $\boldsymbol{d}_i$ between each embedding vector and the attractors as

$$\boldsymbol{d}_i = \boldsymbol{a}_i V，\ i = 1,2,\cdots,N \qquad (6)$$

where $\boldsymbol{d}_i \in \mathbf{R}^{1 \times FT}$. By normalizing the distance，each mask will be estimated by sigmoid function as

$$\hat{\boldsymbol{m}}_i = \mathrm{sigmoid}(\boldsymbol{d}_i) \qquad (7)$$

4）Update the weights of the network by minimizing the reconstruction error as follows：

$$L = \frac{1}{N} \sum_i \| X \odot (\boldsymbol{m}_i - \hat{\boldsymbol{m}}_i) \|_2^2 \qquad (8)$$

### 2.2.3   Testing phase

1）Calculate the phase of spectrogram；

2）Generate the embedding using the previous trained model；

3）Cluster the last embedding using GMM；

4）Find the attractors，which are the center of the clusters；

5）Estimate the mask for each speaker by calculating the distance $\boldsymbol{d}_i$ following Step 3 in the Training Phase；

6）Reconstruct the speech signal for each speaker by multiply the magnitude of the spectrogram by the corresponding estimated mask，and then apply inverse STFT using the phase of the mixture calculated in Step 2.

## 3   Experimental Results

### 3.1   Network Architecture

The network is similar to that of DANet，but differs in the way of extracting embeddings and in the clustering algorithm. The dimension of the input feature is 129. The optimizer algorithm is ADAM with training rate starting at $10^{-3}$, and will be halved if the validation error does not reduce in 3 epochs. The number of epochs was chosen as 150. Four Bi-directional GRU layers were used, and each had 600 units. The dimension of the embedding vector was set to be 20[14], so each T-F bin in spectrogram space was represented by a vector of dimension 20 in embedding space, resulting in the space larger than spectrogram by 20 times. After BGRU layers, there is fully connected feed forward network, which maps from 129 input to 2580 as output dimension（ $20 \times 129 = 2580$ ）.

### 3.2 Dataset

SinceWSJ corpus is not a free dataset, another one was searched. TIMIT is a free speech dataset. A new competitor dataset named TIMIT-2mix was constructed by randomly choosing utterances from different speakers, which were down-sampled to 8 kHz to reduce computation and save memory, and then mixed at various randomly selected SNRs within ［−3 dB, 3 dB］. TIMIT-2mix is similar to WSJ0-2mix[14], and it contains 30 h training set, 10 h validation set, and 5 h testing set. The input feature is logarithm of magnitude of the spectrogram computed by Short-Time Fourier Transform （ STFT ） using 32 ms window length with 75% overlapping between the windows and the square root of Hanning window.

### 3.3 Evaluation Metrics

The proposed model was evaluated using objective measures, which are widely used in speech separation field[19], such as Signal to Distortion Ratio （ SDR ）, Signal to Interference Ratio （ SIR ）, Signal to Artifact Ratio （ SAR ）, and Perceptual Evaluation Speech Quality （ PESQ ） score[20].

### 3.4 Results

#### 3.4.1 Results of complexity decrease

Two models were trained, one using BLSTM and the other using BGRU. The training was performed by a computer with the following specifications: Intel core-i5-8250U central processing unit at 1.8 GHz, GPU NVidia MX130, 8 GB RAM under Windows 10, 64-bit platform, and Python-3.5 under jupyter notebook web-based environment. Table 1 shows the comparison results between the two models using the same dataset TIMIT-2mix.

Table 1 indicates that BGRU model was less complex than BLSTM model, where the improvement in the number of parameters was about 20.7% and 17.9% in training period.

**Table 1　Comparison of the two models in terms of complexity**

| Metrics | Epochs | Dataset | # Parameters | Training period |
|---------|--------|---------|--------------|-----------------|
| BLSTM | 150 | TIMIT-2mix | 9079380 | 186 h 7 min |
| BGRU | 150 | TIMIT-2mix | 7197180 | 152 h 46 min |

Reduction of function error was greater in the proposed model, as shown in Table 2.

**Table 2　Reduction of validation error versus epochs**

| Epoch number | BLSTM error | BGRU error |
|--------------|-------------|------------|
| 1 | 1477.61 | 1542.05 |
| 20 | 380.95 | 380.19 |
| 50 | 260.61 | 249.64 |
| 80 | 244.86 | 211.12 |
| 100 | 238.09 | 209.65 |
| 150 | 240.70 | 212.50 |

It can be seen in Fig.3 that BGRU converged faster than BLSTM.

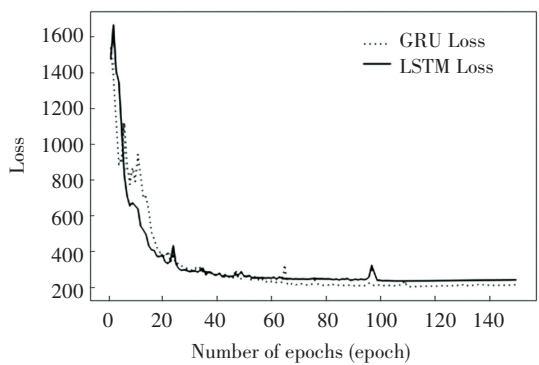

**Fig.3　Comparison of error loss between GRU and LSTM**

#### 3.4.2 Results of accuracy increase

The architecture of the network （ BGRU or BLSTM ） and the clustering algorithm （ k-means or GMM ） can improve accuracy. Effects of the parameters were studied individually and then together.

First, the clustering algorithm was changed to evaluate the effects. Table 3 shows the separation results using BLSTM network with k-means and BLSTM network with GMM algorithm.

**Table 3**  **Comparison of〔BLSTM + k-means, BLSTM + GMM〕in terms of accuracy of separation using TIMIT-2mix dataset**

| Metrics | SDR (dB) | SIR (dB) | SAR (dB) | PESQ |
|---|---|---|---|---|
| BLSTM+k-means | 9.30 | 15.80 | 11.40 | 2.07 |
| BLSTM+GMM | 10.70 | 16.90 | 11.50 | 2.50 |

Then, the architecture of the network was changed to check its effects. Table 4 shows the separation results using BLSTM network with k-means and BGRU network with k-means algorithm.

**Table 4**  **Comparison of〔BLSTM+k-means, BGRU+ k-means〕in terms of accuracy of separation using TIMIT-2mix dataset**

| Metrics | SDR (dB) | SIR (dB) | SAR (dB) | PESQ |
|---|---|---|---|---|
| BLSTM+k-means | 9.30 | 15.80 | 11.40 | 2.07 |
| BGRU+k-means | 11.80 | 18.10 | 12.90 | 2.71 |

Third, the effect of using BGRU with GMM instead of BLSTM with k-means was studied, and the proposed model was compared with the DANet. The two models should be trained by the same dataset for comparison, but WSJ0-2mix is not free. Thus, DANet was trained on TIMIT-2mix, and the DANet model was established, which consists of BLSTM and k-means. All parameters were mentioned in Ref.〔12〕. In this case, the comparison is convincing because the two models were trained using TIMIT-2mix.

As can be seen in Table 5, the proposed model, which depends on BGRU network with GMM, outperformed the DANet model.

**Table 5**  **Comparison of DANet with the proposed model in terms of accuracy of separation using TIMIT-2mix dataset**

| Metrics | SDR (dB) | SIR (dB) | SAR (dB) | PESQ |
|---|---|---|---|---|
| DANet | 9.30 | 15.80 | 11.40 | 2.07 |
| The proposed model | 12.30 | 19.20 | 13.60 | 2.94 |

The results of our model can be seen clearly from Fig.4, where Fig.4(a) and Fig.4(b) show the two separated speakers in time domain and in spectrogram domain, respectively.

The system is language independent, because it can work in Arabic although it learned in English language. It relies on features of the human voice and does not depend on language. According to the study, the separation over Arabic mixture was better in English, for it depends on the speed of talking, and speaking in Arabic is often slower than in English. Table 6 shows the performance of the proposed model on Arabic and English mixtures.

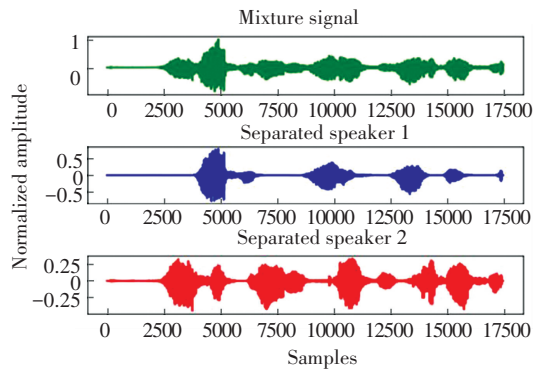

(a) Separation results in normalized time domain (samples)

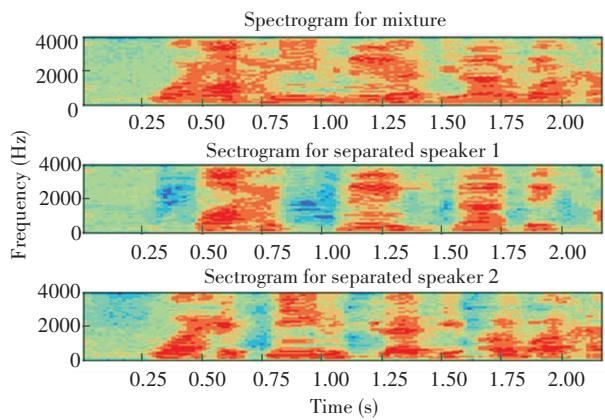

(b) Separation results in spectrogram domain

**Fig.4**  **Speech separation results by the model**

**Table 6**  **Comparison of separation results between Arabic and English mixtures**

| Metrics | SDR (dB) | SIR (dB) | SAR (dB) | PESQ |
|---|---|---|---|---|
| Arabic | 13.20 | 21.00 | 14.80 | 3.11 |
| English | 12.30 | 19.20 | 13.60 | 2.94 |

### 3.5  Discussion

Table 1 reveals how the complexity is reduced by using GRU. GRU is more and more used and is much simpler than LSTM. The reason is that LSTM has three gates and internal memory (cell state), while GRU only has two gates without cell state, which yields to less computational power and faster training. Therefore, GRU can be used to form really deep networks.

As shown in Table 3, the separation accuracy was improved by GMM, and BGRU was more useful in separation speech（Table 4）. In Table 5, the BGRU network with GMM was better than BLSTM with k-means. The architecture of the network and the clustering algorithm had different influences on increasing the accuracy.

In our task（i.e., speech separation）, according to the experimental results, GRU performed better than LSTM, and the reasons for the superiority of GRU over LSTM in speech separation are as follows：

1）GRU does not limit the amount of information added to the cell in each time step, and it is controlled by the forget gate in LSTM, which sometimes leads to the loss of some features.

2）GRU exposes the output（hidden state）in its entirety not as LSTM, which limits the output by hyperbolic tangent and sigmoid functions. It helps the GRU network to find patterns easier.

3）It is possible for GRU to obtain the overall sequence, which makes it more powerful in classification and clustering tasks. While the complexity of LSTM（more gates and cell state）makes it able to learn complicated relationships between words, besides classification and clustering tasks.

In general, the performance of k-means was not better than GMM because of the assumption that all clusters will have spherical models determined by the covariance. In our case, it is not necessary for each cluster to have a spherical shape. GMM is much more flexible in terms of cluster covariance than k-means. Due to the standard deviation parameter, the clusters can take on any shape, rather than being restricted to spheres. k-means is actually a special case of GMM, in which the covariance of each cluster along all dimensions approached zero.

## 4　Conclusions

In this work, a new version of DANet was proposed, which was less complex and more accurate by using BGRU for generating embeddings and GMM with general covariance matrix for each cluster. These modifications in DANet structure improved the speed learning and separation accuracy, which is beneficial in using normal PC for training neural network instead of using high performance PC. It was found that the proposed system was language independent, but it performed better in Arabic than in English. Another advantage of this study is that a new challenging dataset TIMIT-2mix is proposed, which may be an alternative of WSJ0-2mix.

## Acknowledgement

The authors would like to thank Dr. Jonathan Le Roux of Mitsubishi Electric Research Lab for his valuable advices.